\documentclass[prl,twocolumn]{revtex4-1}

\usepackage{amssymb}
\usepackage{natbib}
\usepackage{graphicx}
\usepackage{amsmath}
\usepackage[bookmarks = false]{hyperref}
\usepackage{color}
\usepackage{diagbox}
\usepackage{multirow}

\begin{document}

\title{Semi-Deterministic Entanglement between a Single Photon and an Atomic Ensemble}

\author{Jun Li$^{1,\,2,\,*}$}
\author{Ming-Ti Zhou$^{1,\,2,\,*}$}
\author{Chao-Wei Yang$^{1,\,2}$}
\author{Peng-Fei Sun$^{1,\,2}$}
\author{Jian-Long Liu$^{1,\,2}$}
\author{Xiao-Hui Bao$^{1,\,2}$}
\author{Jian-Wei Pan$^{1,\,2}$}

\affiliation{$^1$Hefei National Laboratory for Physical Sciences at Microscale and Department
of Modern Physics, University of Science and Technology of China, Hefei,
Anhui 230026, China}
\affiliation{$^2$CAS Center for Excellence in Quantum Information and Quantum Physics, University of Science and Technology of China, Hefei, Anhui 230026, China}
\affiliation{$^*$These two authors contributed equally to this work.}

\begin{abstract}
Entanglement between a single photon and a matter qubit is an indispensable resource for quantum repeater and quantum networks. With atomic ensembles, the entanglement creation probability is typically very low to inhibit high-order events. In this paper, we propose and experimentally realize a scheme which creates atom-photon entanglement with an intrinsic efficiency of 50\%. We make use of Rydberg blockade to generate two collective excitations, lying in separate internal states. By introducing the momentum degree of freedom for the excitations, and interfering them via Raman coupling, we entangle the two excitations. Via retrieving one excitation, we create the entanglement between the polarization of a single photon and the momentum of the remaining atomic excitation, with a measured fidelity of 0.901(8). The retrieved optical field is verified to be genuine single photons. The realized entanglement may be employed to create entanglement between two distant nodes in a fully heralded way and with a much higher efficiency.
\end{abstract}

\maketitle

%%%%%%%%%%%%%%%%%%%%%%%%%%%%%%%%%%%%%%%%%%%%%%%%%%%%%%%%%%%%%%%%%%%%%%%%%%%%%%%%%%%%%%%%%%%%%%%%%%%%%%%%%%%%%%%%%%%%%%%%%%%%%%%%%%%%%%%%%%%%%%%%%%%%%%%%%%%%%%%%%%%%%%%%%%%%%%%%%%%%%%%%%%%%%%%%%%%%%%%%%%%%%%%%%%%%%%%%%%%%%%%%%%%%%%%%%%%%%%%%%%%%%
Quantum communication mostly makes use of photons as information carrier~\cite{Gisin2002,Yuan2010}. Scalable extension to ultra long distance~\cite{Briegel1998} and to numerous nodes~\cite{Wallnofer2016} is prohibited due to losses and the probabilistic character of photon sources and detections~\cite{Lvovsky2009}. It is very promising of incorporating matter qubits into quantum communication to solve the scalability issue~\cite{Duan2001a}. In this joint approach, entanglement between a single photon and a matter qubit is an indispensable resource~\cite{Kimble2008a,Afzelius2015}. Multiple pairs of such entanglement may be connected to build a quantum repeater~\cite{Briegel1998} which scales much better than direct transmission of photons through a lossy channel, or to build a quantum network of matter qubits~\cite{Wehner2018,Jing2019} which may enable multiparty quantum communication, synchronization of atomic clocks~\cite{Komar2014}, construction of telescopes with an ultra long baseline~\cite{Gottesman2012}, etc.

Many physical systems~\cite{Simon2010b} have been studied as matter qubits for quantum communication, such as laser cooled ions~\cite{Duan2010} and neutral atoms~\cite{Sangouard2011,Reiserer2015}, impurities~\cite{Gao2015,Atature2018} and rare earth ions~\cite{Tittel2010} in a crystal at cryogenic temperature. To prepare entanglement between a single photon and a matter qubit, it is ubiquitous of using spontaneous emission, which inevitably makes the preparation process probabilistic. In the case of single particles, the emission of photons is typically isotropic and only a very small proportion can be collected into a single-mode fiber~\cite{Hensen2015}. While in the case of an ensemble of particles, the creation probability of a single photon in one specific mode has to be kept very low to limit the contribution of high-order events~\cite{Sangouard2011}. To solve this probabilistic issue, a cavity which couples strongly with a single photon may be employed for the single-particle approaches~\cite{Reiserer2015}. While for an ensemble-based system, a mechanism of nonlinear interaction is required to inhibit high-order excitations. Rydberg blockade is such a mechanism~\cite{Saffman2010c}. Harnessing Rydberg blockade, single collective excitations can be prepared in a deterministic fashion and later retrieved as single photons into a well-defined mode via collective enhancement~\cite{saffman2002creating}. In previous experiments, preparation and manipulation of single~\cite{Dudin2012,Li2013g} and double~\cite{ebert2014atomic,li2016hong} excitations have been demonstrated. Nevertheless, entanglement between a single photon and a Rydberg atomic ensemble is still yet to be realized.

\begin{figure}[h]
\includegraphics[width=0.9\columnwidth]{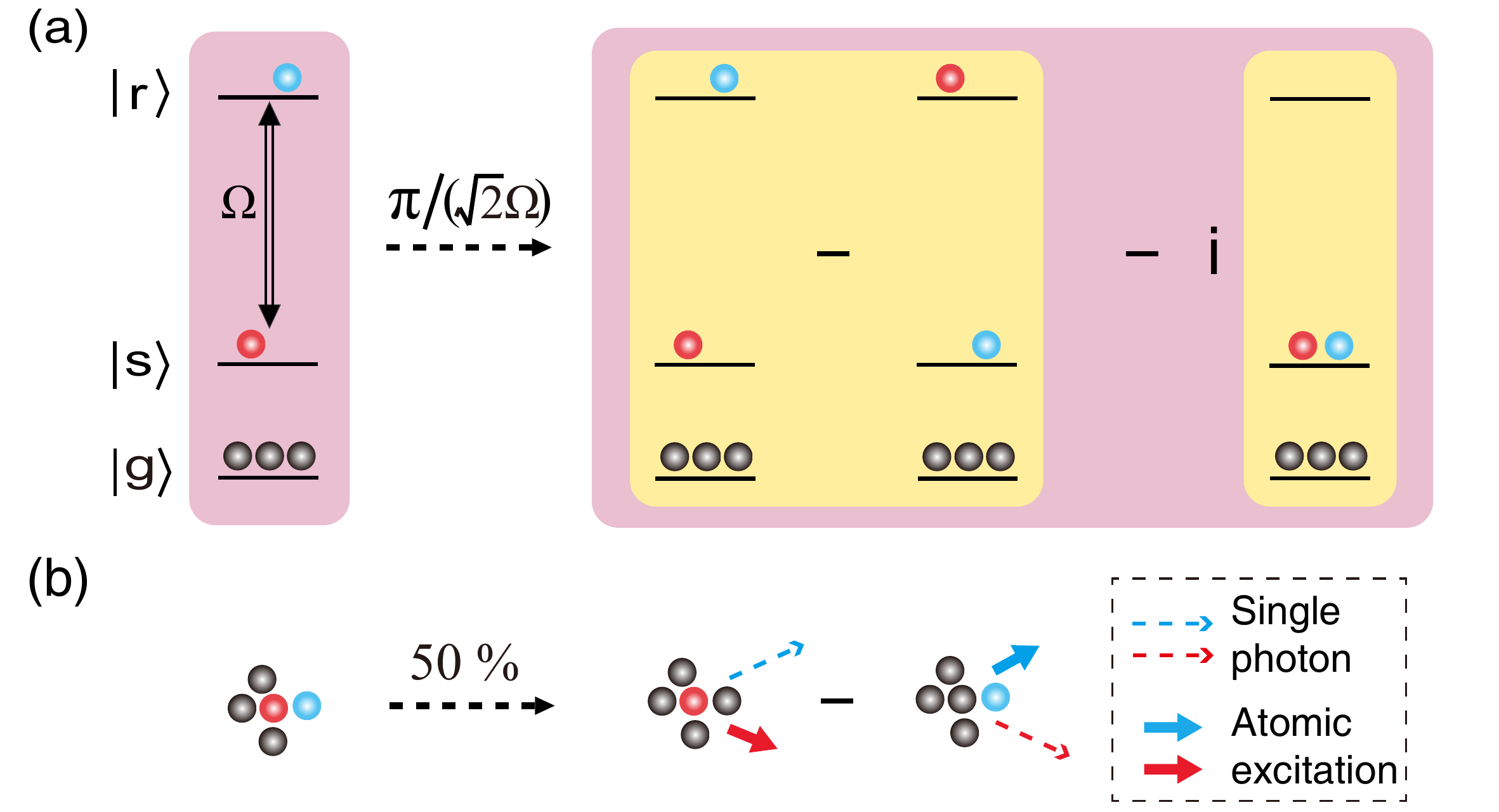}
\caption{Entanglement generation scheme. (a) State evolution of the atomic collective excitation pair after a $\pi$/$\sqrt{2}$ pulse, i.e. ${\rm t}=\pi/(\sqrt{2}\Omega)$, under Rydberg blockade. Red or blue balls represent distinct collective excitations with distinguishable momentum modes. (b) Semi-deterministic atom-photon entanglement generation. By converting the Rydberg excitation into a phase-matched single photon, the momentum of the read-out photon becomes entangled with the momentum of the remaining atomic excitation, with an intrinsic efficiency of 50\%.}\label{concept}
\end{figure}

In this paper, we propose and experimentally realize an atom-photon entanglement scheme which has an intrinsic efficiency of 50\%. Our scheme is illustrated in Fig.~\ref{concept}. We consider a mesoscopic atomic ensemble composed of $N$ neutral atoms in the regime of ensemble blockade. All atoms are initially prepared in a ground state of $|g\rangle$. We make use of Rydberg blockade to prepare two collective excitations, with one in a ground state $|s\rangle$ and another in a Rydberg state $|r\rangle$. The momentums of the two excitations are set to be different, i.e. $\hbar \textbf{k}_{\rm{1}}$ for $|s\rangle$ and $\hbar \textbf{k}_{\rm{2}}$ for $|r\rangle$ respectively (hereafter $\hbar$ is assumed to be 1 for simplicity). Thus, the joint state for the two excitations can be expressed as $|\Psi(0)\rangle=M\Sigma_{jk}^N {\rm{e}}^{i(\textbf{k}_{\rm{1}}\cdotp \textbf{x}_j+\textbf{k}_{\rm{2}}\cdotp \textbf{x}_k)}|s_jr_kg^{N-2}\rangle\equiv|R_{2},S_{1}\rangle$, where $\textbf{x}_{j(k)}$ is the coordinate of the $j(k)$th atom, $M=1/\sqrt{N(N-1)}$ is a normalizing coefficient.{\iffalse and $|S(R)_{1(2)}\rangle$ denotes the collective excitations for simplicity with the subscripts $1$ and $2$ representing their momentums.\fi}~Since all the atoms locate inside the Rydberg blockade radius, the momentum-distinguishable collective excitations can be regarded as two super atoms, and the momentum symbols are just their sequence numbers. Then we apply a driving laser field which resonantly couples the transition $|s\rangle\leftrightarrow|r\rangle$ with a Rabi frequency of $\Omega$, and for simplicity we assume the driving laser induces no momentum kick. Different from noninteracting linear systems \cite{pan2012multiphoton,lester2018measurement}, Rydberg interaction between the super atoms prohibits double Rydberg excitations and it results in a nonlinear process with a $\sqrt{2}$ enhancement~\cite{miroshnychenko2009observation,ebert2014atomic}. At ${\rm t}=\pi/(\sqrt{2}\Omega)$, the collective state evolves into $|\Psi({\rm t})\rangle=(1\sqrt{2})(|\Psi^-\rangle-i|S_{1},S_{2}\rangle)$, where $|\Psi^-\rangle=(1/\sqrt{2})(|R_{2},S_{1}\rangle-|R_{1}, S_{2}\rangle)$ (See Supplemental Material for details). Thus there is a probability of 50\% to obtain $|\Psi^-\rangle$ with two momentum-entangled collective excitations remaining in $|r\rangle$ and $|s\rangle$ respectively. By mapping the Rydberg excitations onto phase-matched single photons, as shown in Fig.~\ref{concept}(b), we can prepare the atom-photon entanglement with an efficiency of 50\%, which we refer to semi-deterministic.

\begin{figure}[h]
\includegraphics[width=\columnwidth]{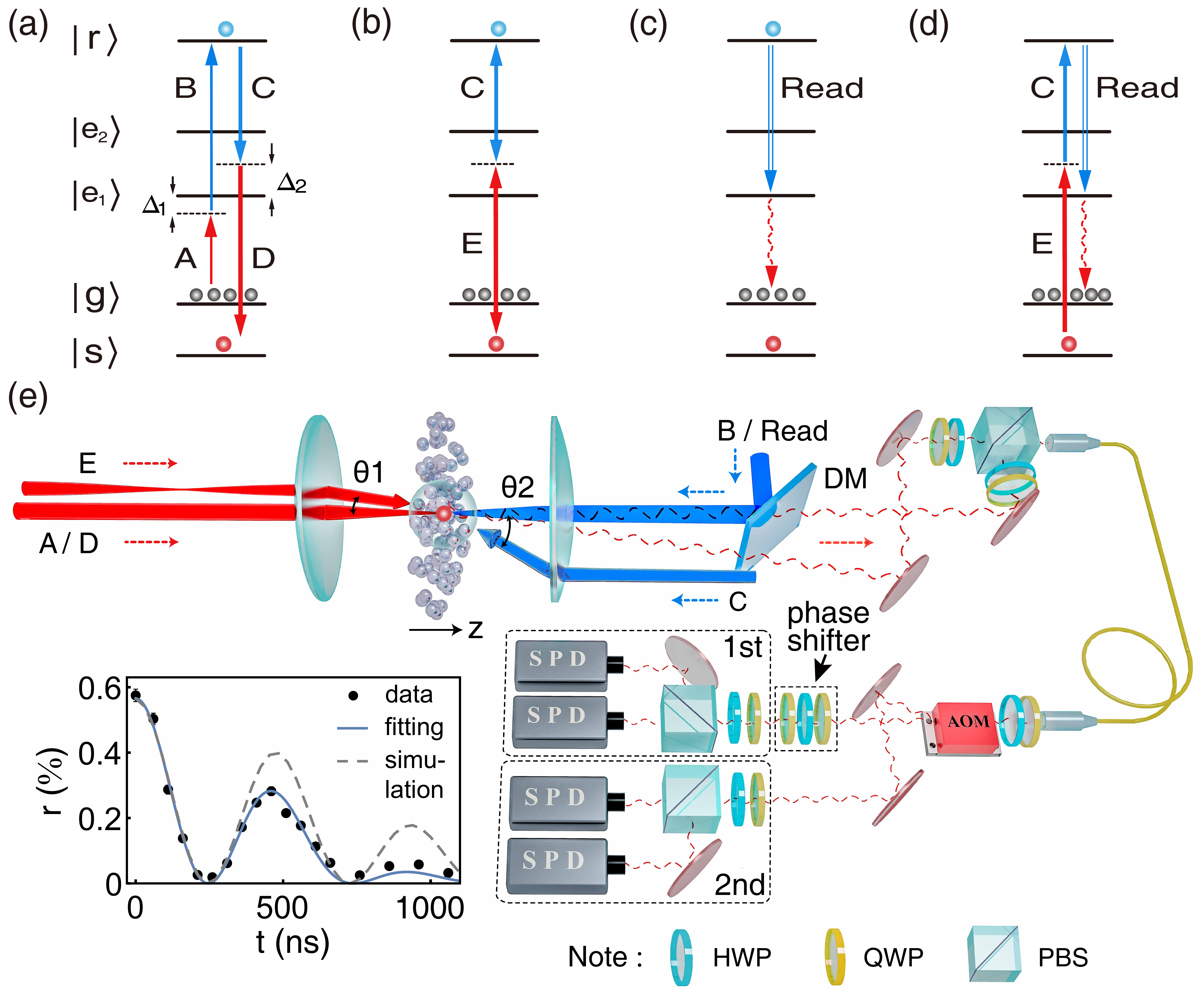}
\caption{(a) Generation of the the two-excitation collective state. The intermediate states are $|e_1\rangle=|5P_{1/2},F^\prime=1,m_{F^\prime}=+1\rangle$ and $|e_2\rangle=|5P_{1/2},F^\prime=2,m_{F^\prime}=+1\rangle$. The single-photon detunings relative to $|e_1\rangle$ ($|\Delta_1|$,  $|\Delta_2|$) are $2\pi\times$(40, 610) MHz respectively, and the single-photon Rabi frequencies of ($A$, $B$, $C$, $D$) coupling $|e_1\rangle$ are $2\pi\times$(3, 6, 33, 23) MHz respectively. The two-photon couplings are all resonant. (b) The Raman coupling. $E$ has nealy the same detuning and Rabi frequency with $D$, but different wave vector. (c) Read-out of the first photon for entanglement generation. We use laser pulses resonant with $|r\rangle$ and $|e_1\rangle$ to read out the Rydberg excitations. (d) Read-out of the second photon for entanglement measurement. The ground-state excitations are also read out through $|r\rangle$ to avoid the noise from the read pulses. (e) The experimental configuration. The thickness of the atomic ensemble is approximately 13 $\mu$m along the z direction defined by a 3 Gauss magnetic field. The beam waists/radii of ($A$, $B$, $C$, $D$, $E$) at the atoms are (7, 7, 13, 520, 520) $\mu$m respectively. $\theta_1$ and $\theta_2$ are about 5$\,^{\circ}$ and 7$\,^{\circ}$ respectively. DM denotes dichroic mirror, HWP denotes half-wave plate, and QWP denotes quarter-wave plate. Inset: Rabi oscillation between $|R_{2}\rangle$ and $|S_{4}\rangle$. We record the photon counting rate of remaining $|R_{2}\rangle$, r, with different Raman pulse durations, t. The fitting oscillation period T is 492(5)~ns, corresponding to a $\pi/\sqrt{2}$ pulse of $\simeq$174~ns. The gray dashed line is the simulation result.}\label{scheme}
\end{figure}

The atomic energy levels and the experimental configuration are depicted in Fig.~\ref{scheme}. Here $|g\rangle=|5S_{1/2},F=2,m_{F}=+2\rangle$ and $|s\rangle=|5S_{1/2},F=1,m_{F}=0\rangle$ are the hyperfine Zeeman sublevels of the ground states of $^{87}\!\rm{Rb}$ atoms, and $|r\rangle$ represents the Rydberg state $\left|81S_{1/2}, m_{J}=+1/2\right>$. The atoms are manipulated within a small volume $\sim$10 $\mu$m in all three dimensions (see Ref.~\cite{li2016hong} and the caption of Fig.~\ref{scheme} for details). Using Rydberg blockade we can deterministically create $|\Psi(0)\rangle$, with the $\pi$ pulse sequences ($A$+$B$)$\to$($C$+$D$)$\to$($A$+$B$) as shown in Fig.~\ref{scheme}(a). As a result, $|S_{1}\rangle$ has a momentum of $\textbf{k}_{\rm{1}} = \textbf{k}_{\rm{A}} + \textbf{k}_{\rm{B}} - \textbf{k}_{\rm{C}} - \textbf{k}_{\rm{D}}$, and $|R_{2}\rangle$ has a momentum of $\textbf{k}_{\rm{2}} = \textbf{k}_{\rm{A}} + \textbf{k}_{\rm{B}}$. Next we apply the Raman beams ($C$+$E$) to couple the two excitations as shown in Fig.~\ref{scheme}(b). Considering the momentum kick of $\Delta\textbf{k} = \textbf{k}_{\rm{C}} + \textbf{k}_{\rm{E}}$ given by the Raman lights, the atomic state after Raman coupling is $|\Psi({\rm t})\rangle'=(1\sqrt{2})(|\Psi^-\rangle'-i|S_{1},S_{4}\rangle)$, with $|\Psi^-\rangle'=(1/\sqrt{2})(|R_{2},S_{1}\rangle-|R_{3}, S_{4}\rangle)$, $\textbf{k}_{\rm{3}} = \textbf{k}_{\rm{1}} + \Delta\textbf{k}$, $\textbf{k}_{\rm{4}} = \textbf{k}_{\rm{2}} - \Delta\textbf{k}$. For the validity of our scheme, one has to ensure $\textbf{k}_{\rm{2}} \neq \textbf{k}_{\rm{3}}$ and $\textbf{k}_{\rm{1}} \neq \textbf{k}_{\rm{4}}$, which is achieved by injecting beam $D$ and $E$ through slightly different directions, $\textbf{k}_{\rm{D}} \neq \textbf{k}_{\rm{E}}$. To generate atom-photon entanglement, we retrieve the Rydberg excitation by applying the read beam as shown in Fig.~\ref{scheme}(c), and the momentum entangled atomic state $|\Psi^-\rangle'$ is converted to $|\Psi\rangle=(1/\sqrt{2})(|\textbf{k}_{\uparrow}\rangle|S_{1}\rangle-|\textbf{k}_{\downarrow}\rangle|S_{4}\rangle)$, where $\textbf{k}_{\uparrow} = \textbf{k}_{\rm{2}} - \textbf{k}_{\rm{Re}}$ and $\textbf{k}_{\downarrow} = \textbf{k}_{\rm{3}} - \textbf{k}_{\rm{Re}}$ denote the momentum states of the read-out photons with $\textbf{k}_{\rm{Re}}$ being the momentum of the read beam. In our experiment, beam $A$ and $D$ are in the same direction, $\textbf{k}_{\rm A} \approx \textbf{k}_{\rm D}$, and the read beam and beam $B$ are in the same direction, $\textbf{k}_{\rm Re} \approx \textbf{k}_{\rm B}$. Thus we have $\textbf{k}_{\uparrow} \approx \textbf{k}_{\rm A}$ and $\textbf{k}_{\downarrow} \approx \textbf{k}_{\rm E}$. After a configurable delay of 300~ns, we retrieve the ground state excitation as well to verify the atom-photon entanglement by applying ($C$+$E$) and the read beam as shown in Fig.~\ref{scheme}(d). Thus the atom-photon entanglement is converted into a photon pair entanglement $|\Psi\rangle_{\rm {mo}}=(1/\sqrt{2})(|\textbf{k}_{\uparrow}\rangle|\textbf{k}_{\downarrow}\rangle-|\textbf{k}_{\downarrow}\rangle|\textbf{k}_{\uparrow}\rangle)$. Note the Rydberg excitation is always the first to be read out. By the use of waveplates and a polarization beam splitter (PBS), as shown in Fig.~\ref{scheme}(e), the momentums of photons can be conveniently transformed as polarizations. For the detection and analysis of the entanglement, we collect the photons with a single mode fiber and use an acousto-optic modulator (AOM) to switch the detection channels between the two photons for independent measurement. A phase shifter between horizontal polarization (H) and vetical polarization (V) is inserted in the path of the first photon, to introduce a variable phase term in the detected photonic state $|\Psi\rangle_{\rm pol}=(1/\sqrt{2})(|{\rm{H}}\rangle|{\rm{V}}\rangle-{\rm{e}}^{i{\rm{\phi}}}|{\rm{V}}\rangle|{\rm{H}}\rangle)$.

We first verify the success of Rydberg blockade by measuring the second-order autocorrelation $g^2$ of the photons converted from the Rydberg excitation $|R_{2}\rangle$. The result $g^2(0)$=0.062(7), which is primarily limited by the noise of the single photon detectors (SPDs), proves the effectiveness of Rydberg blockade and lays the foundation for the following experimental operations. And also we measure the Rabi oscillations during the preparation and manipulation of single collective excitations, with details given in the Supplemental Material. In particular we shows the Rabi oscillation between $|R_{2}\rangle$ and $|S_{4}\rangle$ as an inset of Fig.~\ref{scheme}. In this measurement, a $|R_{2}\rangle$ excitation is first prepared with a 200~ns long exciting pulse, later manipulated by the Raman beams ($C$+$E$) with a tunable duration, and finally retrieved as single photons. Our numerical simulation shows that the damping is mainly due to Rydberg spin-wave dephasing and photon scattering from intermediate states (more details are given in the Supplemental Material).

\begin{figure}[h]
\includegraphics[width=1\columnwidth]{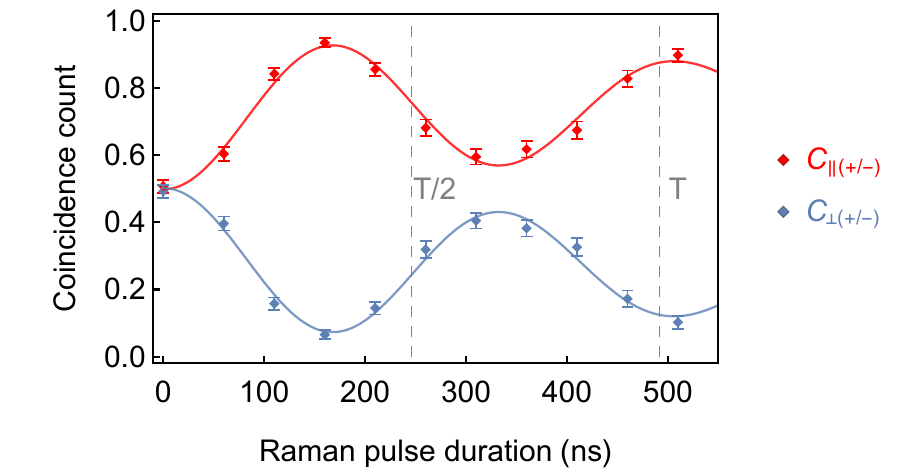}
\caption{Photonic polarization correlations in $+/-$ basis versus Raman pulse duration. The red and the blue diamonds respresent $C_{\rVert}$ and $C_{\bot}$ respectively. We use damped sinusoidal functions to fit the data and get the oscillation period averaged as 339(5)~ns. The dashed lines as labeled correspond to T/2 and T in the single excitation measurement (Fig.~\ref{scheme} inset) where T is the Rabi oscillation period.}\label{duration}
\end{figure}

Next, we measure the dynamics of two excitations under Raman manipulations via polarization correlation measurement of the read-out photons. At $t=0$, the corresponding read-out photonic state should be $|{\rm{H}}\rangle|{\rm{V}}\rangle$, while at $t=\pi/(\sqrt{2}\Omega)$, the corresponding photonic state is $(1/\sqrt{2})(|{\rm{H}}\rangle|{\rm{V}}\rangle+|{\rm{V}}\rangle|{\rm{H}}\rangle)$ (for $\phi=\pi$ in $|\Psi\rangle_{\rm pol}$), and the polarization correlations measured in the $|\pm\rangle=(1/\sqrt{2})(|H\rangle\pm|V\rangle)$ basis will differ radically between the product state and the entangled state. To demonstrate this process, we apply the following pulse sequence: $A$+$B$ for 200~ns, $C$+$D$ for 230~ns, and another $A$+$B$ for 200~ns, to prepare $|\Psi(0)\rangle$. And then variable C+E pulse follows, executing the Raman coupling. In the end we read out each excitation sequentially with a 300~ns interval between them, and record the coincidence counts of all possible polarization pair sets $C_{\rm{++}}$, $C_{\rm{--}}$, $C_{\rm{+-}}$ and $C_{\rm{-+}}$ in the $+/-$ basis. With $\phi\simeq\pi$, the results of coincidence count summations $C_{\rVert(\rm{+/-})}=C_{\rm ++}+C_{\rm --}$ and $C_{\bot(\rm{+/-})}=C_{\rm +-}+C_{\rm -+}$ are shown in Fig.~\ref{duration}, which have been normalized by setting the summations of all coincidence counts as 1 unit. The sinusoidal oscillations reveal the atomic state evolution, and the fitting oscillation period is T$^\prime$~=~339(5)~ns, in good agreement with the theoretical description with a $\sqrt{2}$ factor compared with the single excitation result T= 492(5)~ns. The two-excitation measurement verifies the successful Rydberg blockade betweeen the collective atomic excitaitons.

\begin{figure}[h]
\includegraphics[width=1\columnwidth]{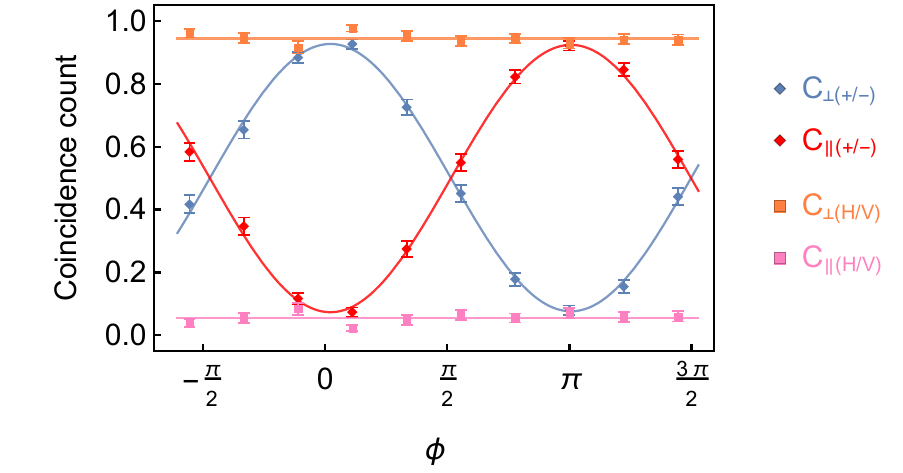}
\caption{Photonic polarization correlations versus $\phi$. In $+/-$ basis $C_{\bot}$ and $C_{\rVert}$ exhibit sinusoidal oscillations in a complementary manner, with a period of $\simeq2\pi$, while in H/V basis $C_{\bot}$ and $C_{\rVert}$ keep as constants approximating to unity and zero respectively.}\label{visibility}
\end{figure}

According to the above result, we set the Raman pulse duration fixed at 170~ns, corresponding to the $\pi/\sqrt{2}$ pulse, and rotate the half-wave plate in the phase shifter, to vary the relative phase $\phi$ in $|\Psi\rangle_{\rm pol}$. it is not difficult to see in H/V basis there should only be $C_{\rm{HV}}$ and $C_{\rm{VH}}$ components, and $C_{\rm{HH}}$ and $C_{\rm{VV}}$ should always be zero, while in $+/-$ basis the results of $C_{\rm{++}}$, $C_{\rm{--}}$, $C_{\rm{+-}}$ and $C_{\rm{-+}}$ depend on the value of $\phi$. The coincidence patterns are shown in Fig.~\ref{visibility}. We define the polarization visibility as $V=|(C_{\bot}-C_{\rVert})/(C_{\bot}+C_{\rVert})|$ at the maximum oscillation points, and therefore we measure them at $\phi=\pi$ in H/V, $+/-$ and circular $\sigma^+$/$\sigma^-$ bases to comprehensively characterize the quality of the entanglement. The results are $V_{\rm (H/V)}=0.897(17)$, $V_{\rm (+/-)}=0.828(19)$ and $V_{\rm (\sigma^+/\sigma^-)}=0.879(19)$ respectively, corresponding to an entanglement fidelity of $F=(1/4)[1+V_{\rm (H/V)}+V_{\rm (+/-)}+V_{\rm (\sigma^+/\sigma^-)}]=0.901(8)$, which is primarily limited by the background noise of the detectors arising from stray lights and dark counts.

Finally, we characterize the single-photon quality of each read-out photon, by analyze their $g^2(0)$ in H/V, $+/-$ and $\sigma^+$/$\sigma^-$ bases respectively. The results are listed in Table~\ref{tab}. Though degraded by lower signal-to-noise ratio compared with single Rydberg excitation measurement, the values here are still well below 1. Averaging the results in the three bases, $g^2(0)$ of each photon are 0.114(15) and 0.120(17) respectively. It proves both read-out fields are genuine single photons, which is very advantageous when implementing quantum repeater and quantum networks using multiple pairs of entanglement. In comparison, the photonic fields of traditional probabilistic source of atom-photon entanglement with atomic ensembles~\cite{Sangouard2011} are generally not single photons, and typically have $g^2(0)$ value around 2. Thus creating entanglement of remote~\cite{Yuan2010} or multiple~\cite{Jing2019} atomic ensembles with probabilistic schemes is usually spoiled with unwanted contributions of high-order events, which makes heralded entanglement creation difficult, while the genuine single-photon feature of our Rydberg-based scheme makes heralded creation of remote entanglement straightforward and enables further extension of node numbers more efficiently~(See Supplemental Material for discussions).

\begin{table}[h]
\caption{Single-photon quality measurement, $g^2(0)$, of each read-out field.}
\begin{tabular*}{\columnwidth}{@{\extracolsep\fill}cccc}\toprule
	Basis&H/V&$+/-$&$\sigma^+/\sigma^-$\\
	\hline
	\multirow{2}{*}{1st}&\multirow{2}{*}{0.090(25)}&\multirow{2}{*}{0.124(24)}&\multirow{2}{*}{0.127(28)}\\&&&\\
	\multirow{2}{*}{2nd}&\multirow{2}{*}{0.129(33)}&\multirow{2}{*}{0.124(28)}&\multirow{2}{*}{0.106(29)}\\&&&\\
	\hline\hline
\end{tabular*}\label{tab}
\end{table}

Albeit the new scheme has many benefits, successful application of it in quantum repeater and quantum networks requires further improvements of overall efficiency and memory lifetime. At present the entanglement detection rate is limited by the overall detection efficiencies in each channel of $\sim$0.2\%, which includes the phase-matching read-out efficiency less than 3\% due to the small optical depth, the transmission and collection efficiency of 50\%, the SPD efficiency of 60\%, the manipulation efficiencies of laser pulses of 65\% $\sim$ 85\%, the effect of spin-wave dephasing, and so on. Much effort is needed to improve the experimental techniques in the future. For instance, by combining the small atomic ensemble with an optical cavity, we can expect the establishment of an effecient photon-atom inetrface~\cite{bao2012efficient}, enhancing the read-out effeciency by an order of magnitude. Replacing the SPDs with state-of-the-art ones, the single photon detection efficiency can exceed 90\%. And with smaller atomic ensembles, better laser systems, as well as new energy configurations~\cite{levine2018high}, the losses during state manipulations will be greatly inhibited, due to more ideal experimental settings. In addition, the entanglement lifetime is limited by the coherence of the ground-state excitation $|S_{4}\rangle$, which is subject to motion-induced dephasing with a lifetime of $\sim$30 $\mu$s at present~\cite{li2016hong}. It will be prolonged markedly to subsecond level~\cite{yang2016efficient} with the use of an optical lattice in the future.

In conclusion, we have proposed and realized a Rydberg-based scheme which creates entanglement between a single photon and a collective atomic excitation by harnessing the momentum degree of freedom. Rydberg blockade is employed to create single collective excitations deterministically and let the excitations interact. The scheme succeeds with an intrinsic efficiency of 50\%, which enables more efficient creation of remote entanglement. The photons created are genuine single photons, which makes heralded creation of remote entanglement straightforward. Limitations in current experimental demonstration can be overcome by incorporating the existing techniques, such as using a ring cavity with moderate finesse to improve the photonic retrieval efficiency~\cite{Yang2015a} and using optically lattice to improve the memory lifetime~\cite{yang2016efficient}. With these developments, our new source of atom-photon entanglement may become a fundamental building block for future quantum repeater and quantum networks~\cite{Wehner2018}.

This work was supported by National Key R\&D Program of China (No. 2017YFA0303902), Anhui Initiative in Quantum Information Technologies, National Natural Science Foundation of China, and the Chinese Academy of Sciences. J. L. acknowledges support from China Postdoctoral Science Foundation.
%%%%%%%%%%%%%%%%%%%%%%%%%%%%%%%%%%%%%%%%%%%%%%%%%%%%%%%%%%%%%%%%%%%%%%%%%%%%%%%%%%%%%%%%%%%%%%%%%%%%%%%%%%%%%%%%%%%%%%%%%%%%%%%%%%%%%%%%%%%%%%%%%%%%%%%%%%%%%%%%%%%%%%%%%%%%%%%%%%%%%%%%%%%%%%%%%%%%%%%%%%%%%%%%%%%%%%%%%%%%%%%%%%%%%%%%%%%%%%%%%%%%%
%merlin.mbs apsrev4-1.bst 2010-07-25 4.21a (PWD, AO, DPC) hacked
%Control: key (0)
%Control: author (8) initials jnrlst
%Control: editor formatted (1) identically to author
%Control: production of article title (-1) disabled
%Control: page (0) single
%Control: year (1) truncated
%Control: production of eprint (0) enabled

%%%%%%%%%%%%%%%%%%%%%%%%%%%%%%%%%%%%%%%%%%%%%%%%%%%%%%%%%%%%%%%%%%%%%%%%%%%%%%%%%%%%%%%%%%%%%%%%%%

\clearpage

\setcounter{figure}{0}
\setcounter{equation}{0}
\renewcommand{\thefigure}{S\arabic{figure}}
\renewcommand{\theequation}{S\arabic{equation}}
\makeatletter

\section{SUPPLEMENTAL MATERIAL}

\subsection{Detailed Derivation of the State Evolution}

For simplicity, we assume the driving laser field will not introduce any additional momentum kick. As it couples $|s\rangle$ and $|r\rangle$ resonantly and large-detuning coupling between $|g\rangle$ and $|r\rangle$ is negligible, we only need to consider the response of the $j$th and the $k$th atoms residing in $|s\rangle$ and $|r\rangle$ respectively. Starting from $|\psi\!{_{jk}}(0)\rangle=|s_jr_k\rangle$, we can write it as $|\psi\!{_{jk}}(0)\rangle=(1/\sqrt{2})(|\psi^-_{jk}\rangle+|\psi^+_{jk}\rangle)$ where $|\psi^\pm_{jk}\rangle=(1/\sqrt{2})(|s_jr_k\rangle\pm|r_js_k\rangle)$. The following derivation apears straightforward as $|\psi^+_{jk}\rangle$ is definitely coupled with $|s_js_k\rangle$ under Rydberg blockade with an enhanced Rabi frequency of $\sqrt{2}\Omega$, while $|\psi^-_{jk}\rangle$ is decoupled from any other states~\cite{smiroshnychenko2009observation}:
\begin{equation}
\begin{aligned}
|\psi\!{_{jk}}(t)\rangle&=\frac{1}{\sqrt{2}}[|\psi^-_{jk}\rangle+{\cos(\dfrac{\sqrt{2}\Omega t}{2}})|\psi^+_{jk}\rangle\\&~~~~-i{\sin(\dfrac{\sqrt{2}\Omega t}{2}})|s_js_k\rangle] .
\end{aligned}\label{equa5}
\end{equation}

With the phase information of individual atoms preserved in the frozen atomic gas, now we can turn to consider the collective state with the two excitations. Recalling $|\Psi(0)\rangle=M\Sigma_{jk}^N {\rm{e}}^{i(\textbf{k}_{\rm{1}}\cdotp \textbf{x}_j+\textbf{k}_{\rm{2}}\cdotp \textbf{x}_k)}|s_jr_kg^{N-2}\rangle=|R_{2},S_{1}\rangle$, we have
\begin{equation}
\begin{aligned}
|\Psi(t)\rangle&=M\Sigma_{jk}^N {\rm{e}}^{i(\textbf{k}_{\rm{1}}\cdotp \textbf{x}_j+\textbf{k}_{\rm{2}}\cdotp \textbf{x}_k)}|\psi\!{_{jk}}(t)\rangle|g^{N-2}\rangle\\&=\frac{1}{\sqrt{2}}[|\Psi^-\rangle+{\cos(\dfrac{\sqrt{2}\Omega t}{2}})|\Psi^+\rangle\\&~~~~-i{\sin(\dfrac{\sqrt{2}\Omega t}{2}})|S_{1},S_{2}\rangle],
\end{aligned}\label{collective}
\end{equation}
with $|\Psi^\pm\rangle=(1/\sqrt{2})(|R_{2},S_{1}\rangle\pm|R_{1}, S_{2}\rangle)$. This is the collective state evolution of the complete system of N atoms, and at ${\rm t^\prime}=\pi/(\sqrt{2}\Omega)$, we obtain $|\Psi({\rm t^\prime})\rangle=(1\sqrt{2})(|\Psi^-\rangle-i|S_{1},S_{2}\rangle)$.

\subsection{Rabi Oscillation between Collective States}

\begin{figure}[h]
	\includegraphics[width=\columnwidth]{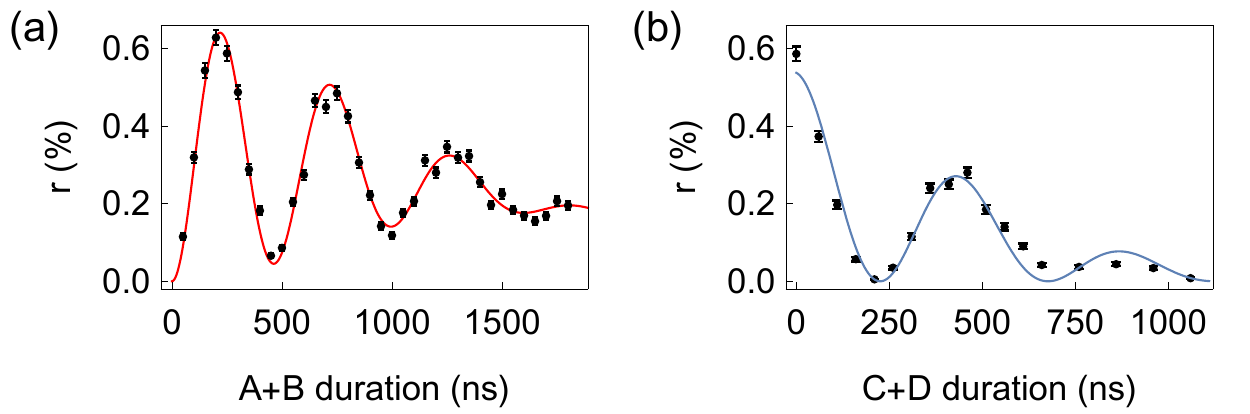}
	\caption{Rabi oscillation between the collective states. (a) Collective Rabi oscillation between $|G\rangle$ and $|R_{2}\rangle$. (b) single-excitation Rabi oscillation between $|R_{2}\rangle$ and $|S_{1}\rangle$. The x axis represents $A$+$B$ or $C$+$D$ pulse duration, and the y axis is the experimental counting rate of the read-out photons converted from $|R_{2}\rangle$. In (b), the initial state is $|R_{2}\rangle$.}\label{s1}
\end{figure}

The collective Rabi oscillation between the initial state $|G\rangle=|g\rangle^N$ and the Rydberg excitation $|R_{2}\rangle$ is shown in Fig.~\ref{s1}(a), by recording the phase-matched single photons converted from the Rydberg excitations. The atom-field coupling is collectively enhanced by a factor of $\sqrt{N_e}$ due to Rydberg blockade, with the effective atom number $N_e\simeq150$. A collective $\pi$ pulse will generate the single excitation in the Rydberg state deterministically. However, the pulse fidelity is experimentally limited by fluctuations of the atom number and the intensities of the laser fields, and the decay of read-out efficiency comes from spin-wave dephasing due to atomic motion and double Rydberg excitations~\cite{dudin2012observation}. Through the envelope of the oscillation fitting, we infer the generation efficiency of the single excitations is better than 80\% with the collective $\pi$ pulse in our experiment.

Fig.~\ref{s1}(b) shows the Rabi oscillation between $|R_{2}\rangle$ and $|S_{1}\rangle$ with $C$+$D$ pulses, which we exploit to prepare the ground-state excitations from the Rydberg ones. Note there is only one atom among $N$ resonantly coupled with the light fields and no collective enhancement in this process. Compared with the collective Rabi oscillation between $|G\rangle$ and $|R_{2}\rangle$, the read-out signal of single-excitation Rabi oscillation between $|R_{2}\rangle$ and $|S_{1}\rangle$ declines much more rapidly. The manipulation losses between the single excitations is a significant factor affecting our experimental efficiency, and thus we clarify the main mechanisms of this damping by numerical simulation, and demonstrate them in the next section below.

\subsection{Sources of Read-Out Losses during Single-Excitation Manipulation}

The Rydberg state $|r\rangle$ and the ground state $|s\rangle$ are coupled by $C$+$D$ (or $C$+$E$) light fields (Fig.~2 in the main text). With the specified intermediate-state detuning $\Delta_2$, the coupling between $|r\rangle$ and $|s\rangle$ is constructively enhanced through $|e_1\rangle$ and  $|e_2\rangle$, and the spatially inhomogeneous light shifts of $C$ are automatically cancelled. However, there are still some sources of Rydberg spin-wave dephasing~\cite{de2018analysis} leading to read-out losses. We use QuTiP~\cite{johansson2012qutip,johansson2013qutip2}, an open-source Python framework for solving the dynamics of open quantum systems, to simulate the single-excitation evolution with various experimental conditions, and try to reveal the causes of the losses.

\begin{figure}[h]
	\includegraphics[width=\columnwidth]{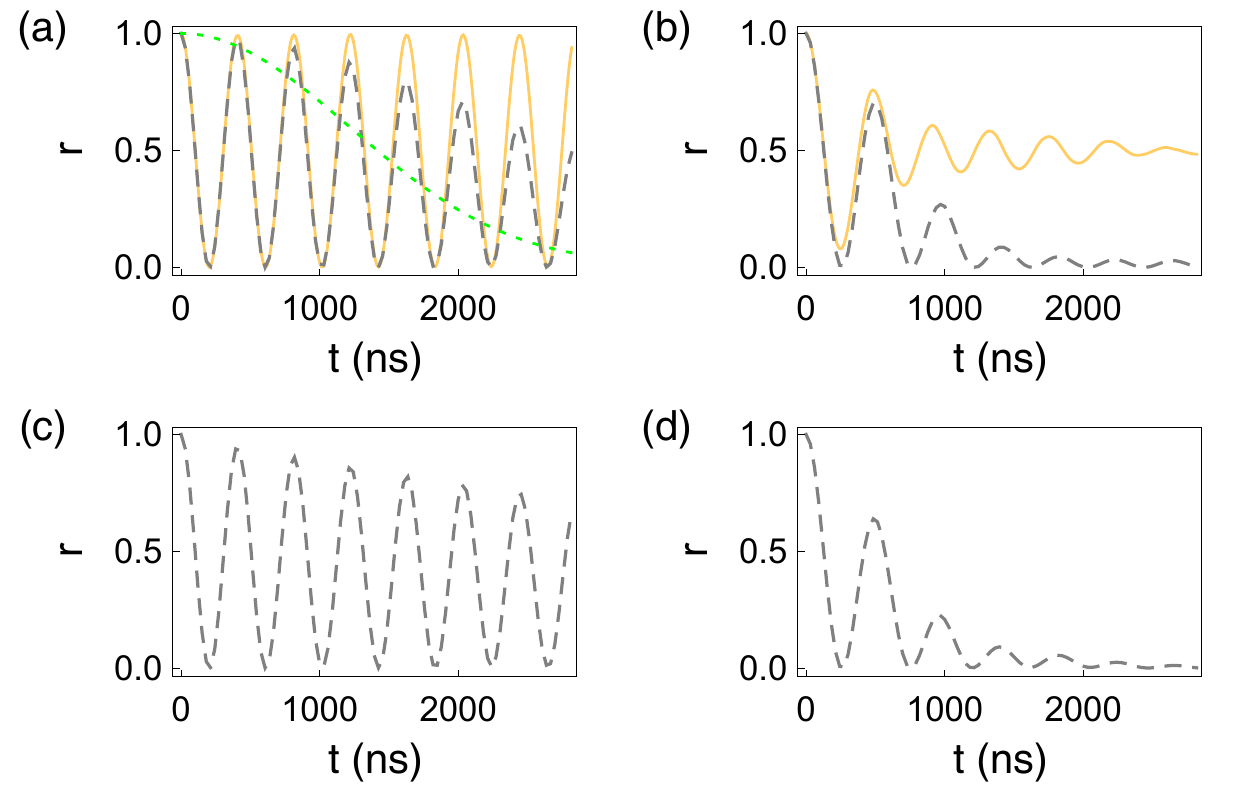}
	\caption{Simulation results. (a) only with the effect of motion-induced Rydberg dephasing. (b) only with the effect of light inhomogeneity. (c) only with the effect of intermediate-state scatterring. (d) with all the factors above. The parameters are set as in the experiment.}\label{s2}
\end{figure}

In the simulations, the propagting directions of the laser beams are set generally the same as in the experiment, while the angles between them are simply ignored. The central Rabi frequencies are kept the same as in the experiment when the beam waists are adjusted for comparison. Other parameters not mentioned all follows the experiment settings, and the initial state is the collective Rydberg excitation, or spin wave, $|R_{2}\rangle$, with a Gaussian spatial mode determined by the 7~$\mu$m focused excitation beams. Fig.~\ref{s2} demonstrates the simulation results with various experimental conditions, where the x axes are the light pulse duration and the y axes are the normalized Rydberg component. The yellow solid lines simply denote the population in $|r\rangle$, while the gray dashed lines denote the state projection onto the collective state $|R_{2}\rangle$. In Fig.~\ref{s2}(a), we consider only the effect of atomic motion, by setting the temperature as 150 $\mu$K and the driving laser beams nealy homogeneous with waists of 500 $\mu$m. We see the Rydberg spin wave dephases gradually, though the probability of finding the atom in the Rydberg state oscilates much better. The spin-wave oscillation has a $1/e^2$ decay time of 3.2 $\mu$s, exactly two times the free evolution lifetime of the Rydberg spin wave without any manipulation light fields (green dotted line). It is consistent with our intuition that the excitation stays in the ground state half of the time during the oscillation with a much longer spin-wave lifetime, and the overall decay time is an average of both. It should be noted that 150 $\mu$K is an effective value, deduced from the measured free Rydberg spin-wave lifetime of 1.6 $\mu$s. There may be some other dephasing machanisms associated with the Rydberg states rather than atomic motion, which results in this lifetime value, such as stray electric fields~\cite{dudin2012strongly}. However, it does not change the conclusion, that Rydberg spin-wave dephasing is one of the sources of read-out losses, and it sets the upper limit as twice the free Rydberg spin-wave lifetime.

In Fig.~\ref{s2}(b), we simulate the system with much smaller beam waist of $C$, 10 $\mu$m, approximately as used in the experiment, and the temperature is set to be zero.
Though the inhomogeneous light shifts of $C$ have been eliminated, the spatial inhomogeneity of Rabi frequency is always an experimental defect, and it smears the oscillation significantly as shown in the figure.

Fig.~\ref{s2}(c) shows the effect of the scattering from the intermediate states. It is an additional source of the manipulation losses, which is not included in the previous two simulations.

Fig.~\ref{s2}(d) brings all the effects together and shows the final simulation result, which is also shown in the inset of Fig.~2 in the main text. From the comparisons above, we conclude that Rydberg-sensitive dephasing, light inhomogeneity, and intermediate-state scaterring are the main sources leading to the manipulation losses, and the light inhomogeneity is the dominant one. And our further experimental measurements with lower Rydberg levels and larger beam waists support this judgement. However, in the  high-level cases, there has to be a trade-off between the homogeneity and the Rabi frequency limited by the laser power. Due to the post selection of the phase-matched read-out photons, the dephasing and losses do not degrade the entanglement fidelity markedly in our experiment. However, high-efficiency light-atom interface is crucial in the application of realistic quantum repeaters. Improvements in laser systems as well as new energy configurations \cite{slevine2018high} will be pursued in our future studies.

\subsection{Advantage of the Semi-Deterministic Atom-Photon Entanglement}

\begin{figure}[h]
	\includegraphics[width=\columnwidth]{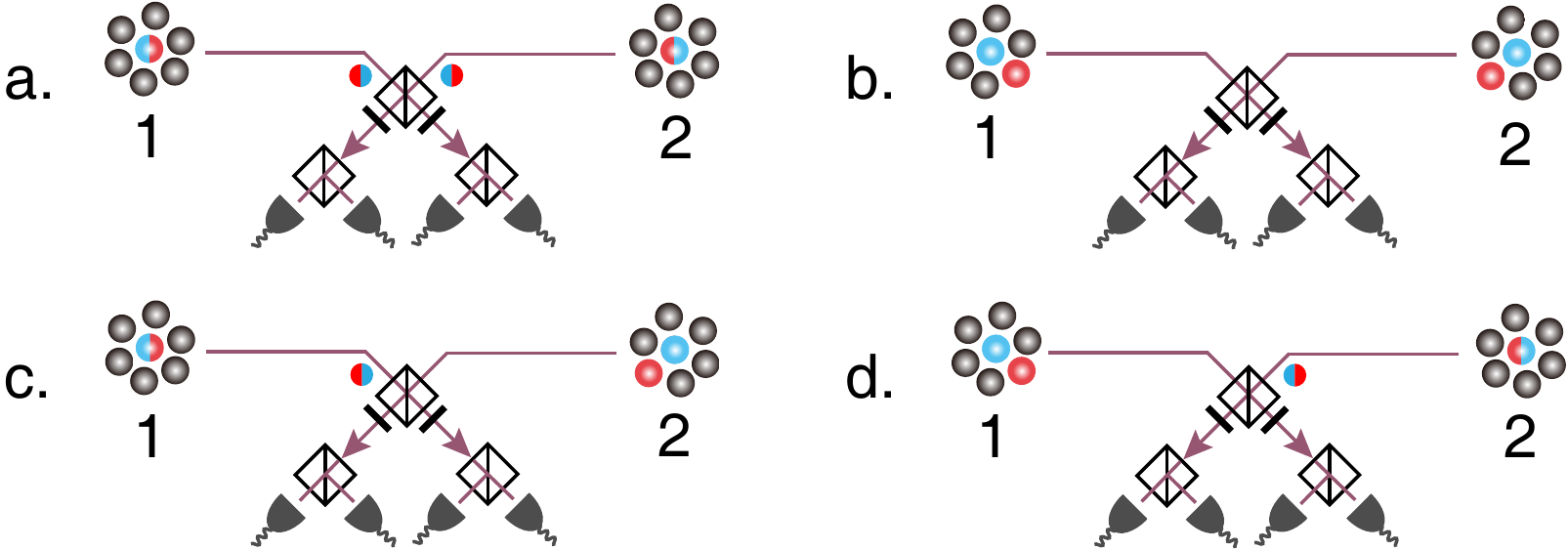}
	\caption{Connecting remote quantum nodes (1 and 2) via entanglement swapping with the entangled atom-photon pairs generated with our scheme. The two-color balls represent entangled atom-photon pairs and the one-color balls only represent atomic excitations with certain momentums. Among the four possible cases $a$, $b$, $c$ and $d$, only case $a$ contributes to the coincidence counts after the Bell-state measurement at the intermediary connection nodes, heralding the establishment of entanglement between 1 and 2.}\label{s3}
\end{figure}

Using the semi-deterministic atom-photon entanglement, we can herald the establishment of entanglement between remote quantum memory nodes without any spurious contributions. The simple fact is illustrated in Fig.~\ref{s3}. For simplicity, we do not consider photon losses in the fibers here. There are four possible cases when connecting the two quantum nodes 1 and 2 by entanglement swapping. Only case $a$, with both of the atom-photon entanglement pairs generated successfully, can obtain the expected two-photon coincidence counts in the middle. And then the remote entanglement between nodes 1 and 2 is prepared successfully for further extensions. In practice, single photons in long-distance fibers may well be absorbed before arriving at the detectors. But the photon losses or nonunit detection efficiency will only bring down the coincidence count rate but can never lead to spurious heraldings. In contrast, due to the existence of sencond-order excitations, the DLCZ-based protocols can not avoid generation of complex states that are not maximally entangled~\cite{zhao2007robust,chen2007fault,jiang2007fast,sangouard2008robust}. And thus the measurement results at the intermediary connection nodes can not ensure the generation of remote entanglement after the first-level entanglement swapping. And following attempts on higher-level entanglement swapping with more classical communication and feedbacks may prove to be futile. As a result, the DLCZ-based quantum repeaters work in a low efficiency.

\end{document}